\documentclass[%
 aip,
 amsmath,amssymb,
 reprint,%
]{revtex4-1}

\usepackage{graphicx}
\usepackage{dcolumn}
\usepackage{bm}

\usepackage[utf8]{inputenc}
\usepackage[T1]{fontenc}
\usepackage{mathptmx}

\DeclareMathOperator{\FCriteria}{||\overline{\Delta\phi}||}

\begin{document}

\preprint{AIP/123-QED}

\title[The rigorous criteria for the phase transitions in Coulomb crystals]{The rigorous criteria for the phase transitions in Coulomb crystals}

\author{A.V. Romanova}
 \email{anromanna98@gmail.com}
\author{S.S. Rudyi}
\author{Y.V. Rozhdestvensky}
\affiliation{%
$^{1}$ITMO University, 49, Kronverkskiy Prospekt, Russia, 197101
}

\date{\today}

\begin{abstract}
The present work suggests rigorous criteria to determine phase transitions in Coulomb crystals in a linear ion trap. The proposed method is based on the analysis of a cross size $\rho_i$ and relative polar angle between neighboring particles $\Delta\phi_i$ as functions of the system parameters such as number of ions, mass and charge of ions, and trap geometry. The analytical interpretation of the phase transitions relies on the analysis of the cross size $\rho$ and a metric $\Phi$ dependent on the norm of the vector $\FCriteria$. We further demonstrate an analysis procedure for the numerical determination of extremes of interpolated functions $\rho$ and $\Phi$. 1D-2D and 2D-3D phase transitions points are defined as points of the greatest growth of functions $\rho$ and $\Phi$, respectively. 
\end{abstract}

\maketitle

\section{\label{sec:level1}Introduction\protect}

Coulomb crystals are unique ordered structures that form in ion traps when ions are cooled below the 10 mK temperature~\cite{drewsen2015ion}. These structures are promising for the simulations of inaccessible systems such as the surface of neutron stars~\cite{pyka2013topological}. Moreover, linear ion crystals, which are also called ion chains, are prospective for the implementation of the quantum computer based on the ion traps~\cite{johanning2016isospaced}.

To describe the dynamics of the linear ion structures, the linear chain oscillation formalism is used~\cite{pedregosa2020defect}, as it considers the interactions only between the nearest neighboring particles by analogy with Toda chains~\cite{toda1967vibration, kotkin2013collection}. In this case, the criterion for the determination of phase transitions is the change of the ratio of the transverse and axial frequencies~\cite{kotkin2013collection, pedregosa2020defect}. Nonetheless, the concept of ``axial frequency'' is associated with a cylindrical effective potential model, which causes a linear restoring force in quadrupole traps~\cite{gerlich1992inhomogeneous}. In the case of a real system, the shape of the effective potential can differ significantly from a parabolic well, for example, in the presence of end-cap electrodes. Toda chains allow to qualitatively describe the linear-zigzag transition in the \textit{ideal quadrupole fields}~\cite{gerlich1992inhomogeneous}. However, this method cannot describe the ``zigzag -- 3D structure'' transition and transitions in a highly deformed effective potential. Thus, the need for an alternative description of Coulomb crystals arises. 

From a physical standpoint, phase transitions result from changes in the total system energy. The total system energy is dependent on parameters of the system such as number of ions, trap geometry, and the elemental composition of the ion crystal. In the general case, finding an analytical solution for the equations of motion is complicated.

On the other hand, the topological dimension of the crystal sharply changes with phase transitions occuring. Each stable configuration can be characterized by geometric parameters -- a cross size and a relative polar angle between neighboring particles. The cross size is the maximum deviation from the symmetry axis $z$, and the relative polar angle is a difference between polar angles of particles with coordinates $z_{i}$ and $z_{i+1}$. In these terms, for the ion chain, the topological dimension equals 1, and the cross size is limited by thermal vibrations. In the case of the zigzag structure, the cross size is greater than thermal vibrations of ions, and relative angles possess values of $-\pi$ and $\pi$. It means that all units of the crystal belong to the same plane, and the topological dimension equals 2. For 3D structures the cross size is also greater than thermal vibrations, and relative angles belong to set $\left[0;2\pi\right]$. Thus, changes in geometric parameters of Coulomb crystals correspond to phase transitions. 
 
In the present work, we propose a rigorous criteria for the phase transitions determination in Coulomb crystals.  We consider a model problem where distances between particles are fixed along $z-$axis. We show that the determination of the phase transitions points reduces to $\rho$ and $\Phi$ analysis, where $\rho$ is a cross size function and $\Phi$ is a metric, which is dependent on the norm of a relative angle vector as $1/\pi\left(\FCriteria-\sqrt{N-1}\right)$, where $N$ is a number of ions. Further, we consider a model problem where Coulomb interaction coefficient increases. For this system, we define linear, zigzag, and three-dimensional structures in terms of $\rho$ and $\Phi$. The approach for the numerical determination of the extrema of interpolated functions $\rho$ and $\Phi$ is proposed.

\section{The charged particles trapping}

For a single particle, the field of the hyperbolic (power) electrodes provides a radial confinement. For this field we generally use the form~\cite{douglas2005linear}
\begin{equation}
  U_1 (x, y)=\frac{e \left[U_0+V \cos{(\Omega t)}\right]}{r^2_0} (x^2-y^2)
  \label{(eq:PotentialEnergyInQuadrupole)}
\end{equation}
where $e$ is the charge of the particle, $U_0$ is the amplitude of the DC voltage applied to the hyperbolic electrodes, $V$, $\Omega$ are the amplitude and the frequency of the AC voltage, respectively, $r_0$ is the trap radius.

\begin{figure}[ht]
		\center{\includegraphics[width=0.9\linewidth]{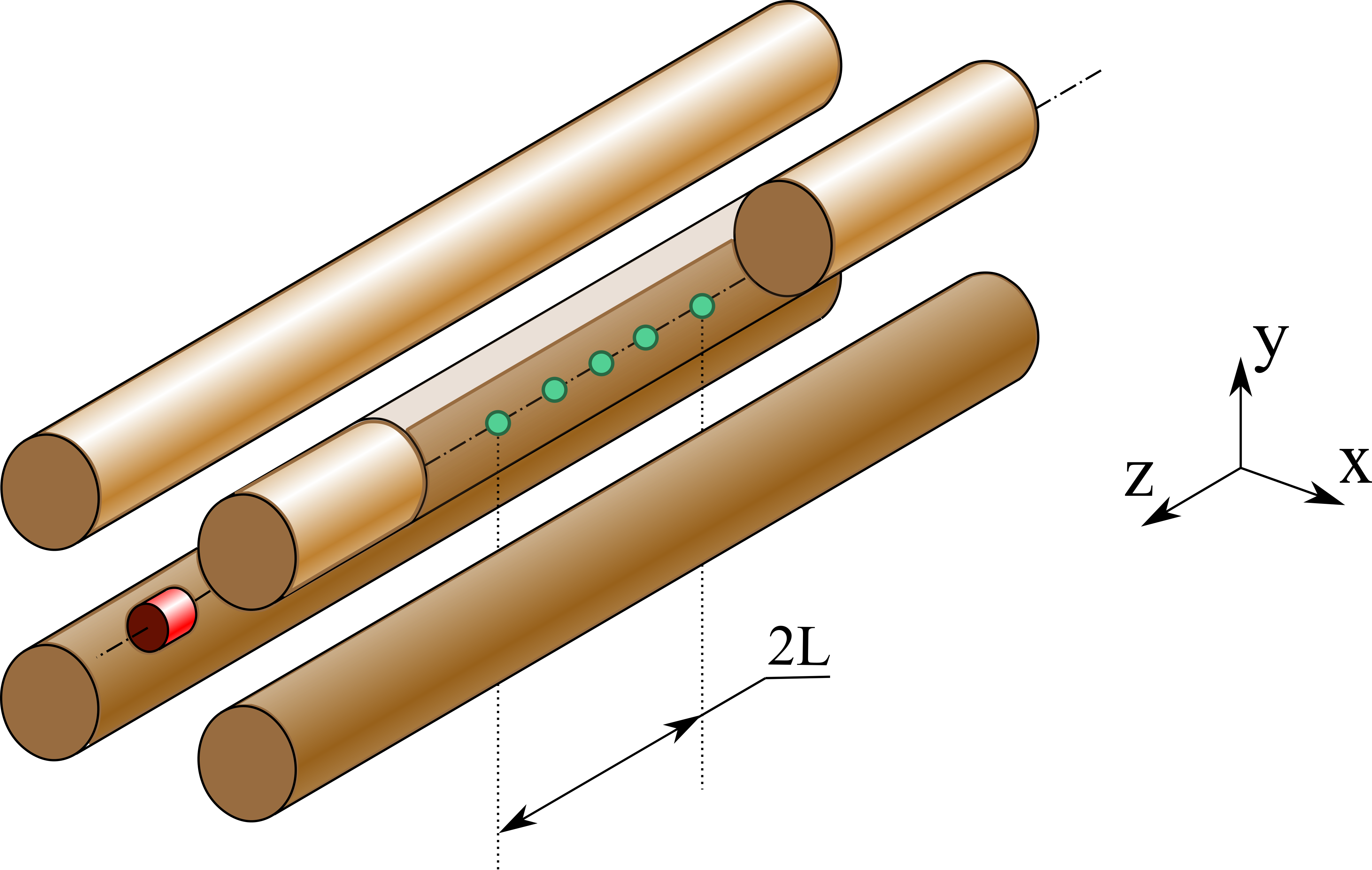} }
	\caption{Schematics of the quadrupole linear trap with four power and two end-cap electrodes, $2L$ is the length of the ion chain at equilibrium}
	\label{img:TrapEndcap}  
\end{figure}

If a trap comprises only power electrodes,  the particle motion is limited along axes $x$ and $y$, but not along axis $z$. In real linear traps, the axial confinement arises from the interaction of the particle with the field of end-cap electrodes $U_\mathrm{end}(x,y,z)$. Figure~\ref{img:TrapEndcap} shows a linear trap with two flat end-cap electrodes.

The total potential energy of $N$ trapped particles is the superposition of the interaction of each particle with fields of power and end-cap electrodes and Coulomb interactions between all trapped particles. The trapped ions hold the local equilibrium position when the temperature is below 10 mK~\cite{drewsen2015ion}. In the presence of damping force, ions form unique ordered structures called \textit{Coulomb crystals}.

One can distinguish one-, two-, and three-dimensional structures. One-dimensional structures are linear Coulomb crystals, which are also called ion string or ion chain~\cite{johanning2016isospaced}. In the case of the one-dimensional structure, equilibrium positions of ions lie along one axis, and the topological dimension equals 1. Two-dimensional Coulomb crystals are called zigzag chains or simply zigzag~\cite{retzker2008double}. In this case, equilibrium positions of ions belong to one plane, and the topological dimension equals 2. In the case of three-dimensional structures, equilibrium positions of ions lie in three-dimensional space, and topological dimension equals 3. When the potential energy changes, one can observe phase transitions, which are attendant with a sharp change in the topological dimension of the Coulomb crystal~\cite{pyka2013topological}.

\section{Model problem with fixed axial ion coordinates}

Let us consider the simplest case of linear Coulomb crystal. In this model problem, equilibrium positions of ions lie along the $z$ axis. In the limit of a large number of ions, the interparticle spacing $\alpha=z_{i+1}-z_i$ is a smooth function of the position, and it is inversely proportional to the density of ions per unit length $n_L$~\cite{morigi2004dynamics}
\begin{equation}
  n_\mathrm{L}(z)=\frac{1}{\alpha(z)}
  \label{eq:DensityAlphaDependence}
\end{equation}
The density of charges for unit length can be evaluated by applying the Gauss theorem to a continuous distribution of charges, which are assumed to be uniformly distributed in an elongated ellipsoid. The resulting one-dimensional density is~\cite{dubin1997minimum}
\begin{equation}
  n_\mathrm{L}=\frac{3}{4} \frac{N}{L} \left(1-\frac{z^2}{L^2}\right)
  \label{eq:Density}
\end{equation}
which is defined for $|z| \ll L$, where $2L$ is the axial length of the string at equilibrium, $N$ is the number of ions. The density~(\ref{eq:Density}) is the leading term in the expansion in powers of $1/\ln{N}$, and it gives a good estimate of the charge distribution in the center of the chain for $N$ sufficiently large. The length is evaluated by minimizing the energy of the crystal and at the leading order in $\ln{N}$ fulfills the relation~\cite{morigi2004dynamics} 
\begin{equation}
  L^3(N)=3 \frac{ke^2}{m \nu^2} N \ln N
  \label{eq:LengthOfTheChain}
\end{equation}
Sequentially substituting Eq.~(\ref{eq:LengthOfTheChain}) into Eq.~(\ref{eq:Density})  and Eq.~(\ref{eq:Density})  into Eq.~(\ref{eq:DensityAlphaDependence}), one finds
\begin{equation}
  \alpha(z)=\frac{4}{3} \frac{(3 ke^2 N \ln N)^{3/2}}{N^3 \sqrt[3]{m_i \nu^2}} \left(\frac{1}{\sqrt[3]{3 ke^2 N \ln N}-z_i^2 \sqrt[3]{m_i \nu^2}}\right)
  \label{eq:AlphaZ}
\end{equation}

Thus, the positions of $i+1$ and $i$ ions are linked by the recurrence relation. If we assume that the position of the first particle is $z_1=0$, then the position of the second particle is given by
\begin{equation}
    z_2=z_1+\alpha (z_1)=0+\alpha (0)=\sqrt[3]{\frac{64m\nu^2}{81e^2kN^4\ln N}}
\end{equation}
Axial positions for the next particles will be $z_3  = z_2  + \alpha (z_2)=\alpha(0)+\alpha(\alpha(0))$, $\ldots$, $z_{i+1}=z_i+\alpha (z_i)$. Thus, the position of the particles is defined by the recurrence relation that depends only on $\alpha(0)$.

The field of end-cap electrodes has no significant effect on the radial motion of particles when voltages applied to the end-cap electrodes are far less than voltages applied to the power electrodes. Consequently, we can consider a simplified model -- the system of material points with charges $e_i$ and masses $m_i$. The positions of material points are fixed along $z$ axis, and we take into account all interactions between these points. Radial confinement in this case is provided by the linear restoring force with the proportionality coefficient $K$ (cylindrical pseudopotential model)~\cite{gerlich1992inhomogeneous}, and $K_x=K_y=K=2\cdot (e^2 V^2)/(m^2 \Omega^4 r_0^4)$. Positions along $z$ axis can be found by using Eq.~(\ref{eq:AlphaZ}). 

Within this approximation, we consider the system of $N$ particles with equal masses $m_1= m_2=\ldots= m_i=m$ and equal charges $e_1=e_2=\ldots=e_i=e$. The particles are trapped by the cylindrical pseudopotential in the ion trap in $(x,y)$ plane. The equations of motion of the $i$-th particle have the form, and the gravity force is neglected here
\begin{equation}
    \frac{d^2 x_i}{d\tau^2} = - A  x_i + B \sum_{i=1}^N \sum_{i\neq j} \left[( x_i- x_j)\kappa(i,j)^{-1}\right] -\frac{d x_i}{d\tau}
   \label{(eq:MotionOX)}
\end{equation}
\begin{equation}
    \frac{d^2 y_i}{d\tau^2} =  - A y_i + B \sum_{i=1}^N \sum_{i\neq j} \left[(y_i- y_j)\kappa(i,j)^{-1}\right] -\frac{d y_i}{d\tau}
   \label{(eq:MotionOY)}
\end{equation}
where $\kappa(i,j)={\left[(x_i- x_j)^2+( y_i- y_j)^2+( z_i- z_j)^2\right]^{3/2}}$, $A = (m \nu)^2/\beta^2$, $B = m k e^2/\beta^2$, $\nu^2=K/m$, $\tau = t \beta/m$, $k=1/(4 \pi \varepsilon_0)$, $\varepsilon_0$ is a dielectric constant, $e$ is the charge of $i$ and $j$ particles, $\beta$ is the effective friction coefficient~\cite{fishman2008structural}. Finally, substituting  the variables A and B into~(6), we obtain the following
\begin{equation}
    \alpha(0)=\sqrt[3]{\frac{64A}{81 B N^4\ln N}}
\end{equation}
Thus, the dynamical system depends only on three parameters -- A, B, and the number of ions, N.

\section{Conditions of phase transitions in Coulomb crystals}
The dynamic system~(\ref{(eq:MotionOX)})--(\ref{(eq:MotionOY)}) is dissipative. Thus, particles aspire to occupy isolated local minima of potential energy. When $\tau\to\infty$, one can observe a formation of \textit{stable configurations}. The formation of a certain stable configuration depends on the system parameters as well as on the initial conditions (the initial coordinates $q_{n_i}[0]$ and velocities $\dot{q}_{n_i}[0]$).

Within the above model problem, we can describe the stable configurations by two geometric parameters. These geometric parameters are the cross size of the Coulomb crystal and relative polar angle. The cross size of Coulomb crystal is a maximum deviation of an ion from the symmetry axis
\begin{equation}
    \rho=\max_{i=1}^{N}{\left(\sqrt{x_i^2+y_i^2}\right)}=\max_{i=1}^{N}{\rho_i}
    \label{eq:rho}
\end{equation}
The relative polar angle is the difference between polar angles $\Delta\phi$ of two particles with $z_i$ and $z_{i+1}$ coordinates. $N-1$ vector, which completely describes a system of $N$ particles, can be defined as
\begin{eqnarray}
    \nonumber\overline{\Delta\phi}=(\Delta\phi_1,\Delta\phi_2,...,\Delta\phi_i,...,\Delta\phi_{N-1})=\phantom{------>}\\
    =(\arctan{\frac{y_1}{x_1}}-\arctan{\frac{y_2}{x_2}},...,\arctan{\frac{y_{N-1}}{x_{N-1}}}-\arctan{\frac{y_{N}}{x_{N}}})
    \label{eq:DeltaPhi}
\end{eqnarray}
The scalar describing a vector $\overline{\Delta\phi}$ is the norm of this vector
\begin{equation}
   \FCriteria=\sqrt{\sum_{i=1}^{N-1}(\Delta\phi_i)^2}
\end{equation}
On the one hand, phase transitions are attended by a sharp change in the relative angles and cross size of the Coulomb crystal. On the other hand, phase transitions occur as a result of a change in the potential system energy. In terms of equations of motion~(\ref{(eq:MotionOX)})--(\ref{(eq:MotionOY)}), the energy of interaction with the trapping potential is proportional to A, and the energy of Coulomb interaction is proportional to B. Thus, $\rho$ and $\FCriteria$ are functions of the parameters of the system $\rho(A,B,q_{n_i}[0],\dot{q}_{n_i}[0])$, $\FCriteria(A,B,q_{n_i}[0],\dot{q}_{n_i}[0])$, where $q_{n_{i}}[0],\dot{q}_{n_{i}}[0]$ are initial conditions of generalized coordinates for $i\in[1,N]$. Each of the stable configurations can be characterized with certain values of $\rho$ and $\FCriteria$. Points of the phase transitions are the points where the functions $\rho$ and $\FCriteria$ have the fastest growth rate. 

\begin{figure}[ht]
	\begin{minipage}[ht]{0.5\linewidth}
		\center{\includegraphics[width=\linewidth]{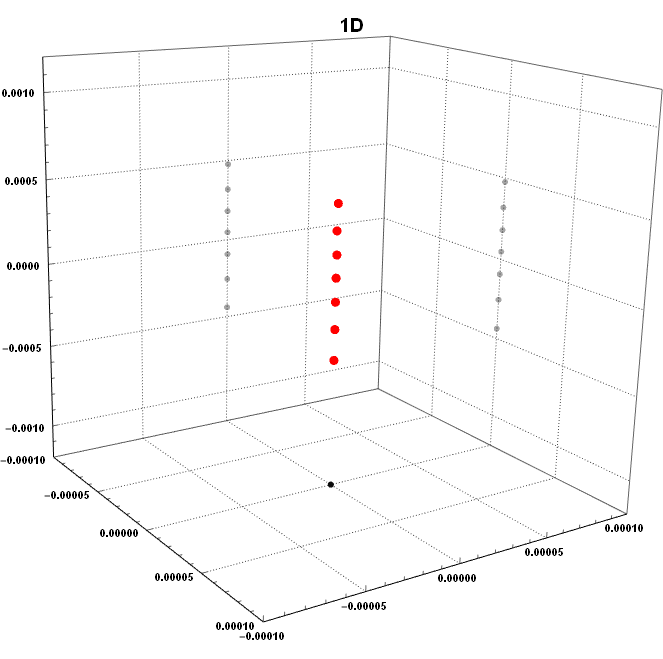}\\(a)}
	\end{minipage}
	\hfill
	\begin{minipage}[ht]{0.5\linewidth}
		\center{\includegraphics[width=\linewidth]{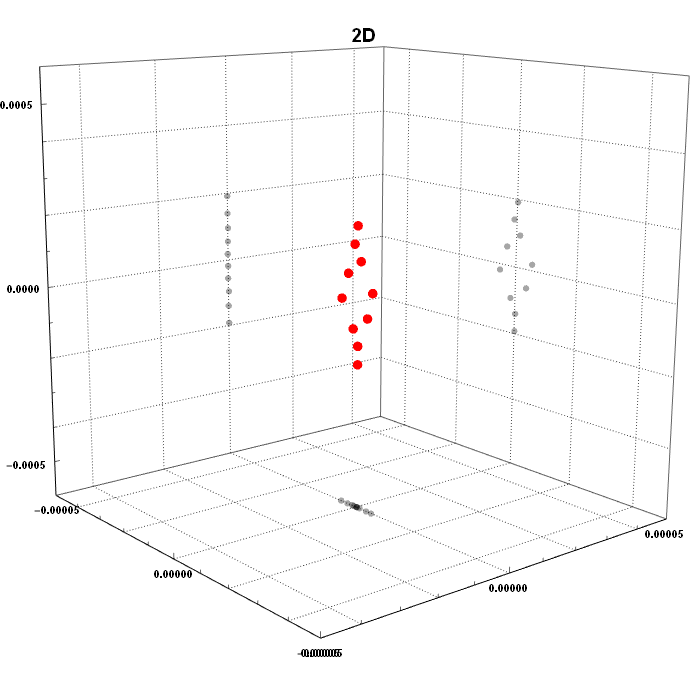}\\(b)}
	\end{minipage}
	\caption{Simulation of stable configurations in the linear trap a) linear ion crystal in the case of $N=7$ particles; b) zigzag in the case of $N=10$ particles.
	\label{Fig:TheLargeImageWithFourParts}}  
\end{figure}

To prove this statement, we consider the phase transitions when the parameter $A$ is fixed. Figure~\ref{Fig:TheLargeImageWithFourParts}a shows the linear ion crystal when $B \ll 1$. In this case, the cross radius $\rho\to0$, and all particles lie on axis $z$. The vector $\overline{\Delta\phi}$ is indeterminate for a linear chain since $(x_i,y_i)=(0,0)$. By this one can see that the linear structure is fully described by the cross size function $\rho(B)$. When the parameter $B$ increases, $\rho$ also increases. Moreover, the function $\rho(B)$ has a cusp at the point $B_{crit}$. Thus, $B_{crit}$ is a point of the phase transition ``string-zigzag''.

We can describe the zigzag configuration with two parameters. Ideally, the relative angles $\Delta\phi_i$ can take discrete values $-\pi$ and $\pi$ (Figure~\ref{Fig:TheLargeImageWithFourParts}b). The norm of the relative angle vector $||\overline{\Delta\phi}||$ takes the form
\begin{equation}
    ||\overline{\Delta\phi}||=||-\pi,\pi,-\pi...||=\sqrt{\sum_{i=1}^{N-1}{\left[(-1)^i\pi\right]^2}}=\pi\sqrt{N-1}
\end{equation}
The subsequent increase of parameter $B$ in the equations of motion~(\ref{(eq:MotionOX)})--(\ref{(eq:MotionOY)}) results in the growth of $||\overline{\Delta\phi}||$. The function $||\overline{\Delta\phi}||$ has a cusp at the point $B_{crit2}$.
Hence, the point of phase transition ``zigzag--three-dimensional structure'' is the minimal value of $B$, at which $||\overline{\Delta\phi}||-\pi\sqrt{N-1}$ is $(0;2\pi]$.

At this point, we can define all stable structures in the considered model as follows\\
1) If for each ion with coordinates $(x_i,y_i)$ the conditions 
\begin{equation}
\rho(B)\leq\varepsilon, \quad\lim_{t\to\infty}{\varepsilon}=0
\label{eq:firstcond}
\end{equation}
are satisfied, then this structure is called \textit{linear structure}.\\ 
2) If for each ion with coordinates $(x_i,y_i)$ the conditions 
\begin{equation}
\rho(B)>\varepsilon,\quad\Phi(B)=\left|\frac{1}{\pi}||{(\overline{\Delta\phi}(B))}||-\sqrt{N-1}\right|\leq\xi,\quad  \lim_{t\to\infty}{\xi}=0
\label{eq:secondcond}
\end{equation}
are satisfied, then this structure is called \textit{zigzag}\\
3) If for each ion with coordinates $(x_i,y_i)$ the conditions 
\begin{equation}
\rho(B)>\varepsilon,\quad\Phi(B)>\xi
\label{eq:thirdcond}
\end{equation}
are satisfied, then this structure is called \textit{three-dimensional structure}. 

\section{Special aspects of the numerical analysis of phase transitions}

We can apply the definitions (\ref{eq:firstcond})--(\ref{eq:thirdcond}) to the models similar to~(\ref{(eq:MotionOX)})--(\ref{(eq:MotionOY)}) when $T\to\infty$. Within these models, we neglect the radiofrequency heating, stochastic effects, which result from the ion re-emission during the laser cooling, and the detection aspects. However, we cannot neglect these effects in the case of real systems. Therefore, $\rho(B)$ and $\Phi(B)$ from the definitions~(\ref{eq:firstcond})--(\ref{eq:thirdcond}) always have nonzero values. It means that $\varepsilon$ and $\xi$ are limited from below by nonzero $\varepsilon_{0}$ and $\xi_{0}$, respectively. The points of the ``linear--zigzag'' and ``zigzag--three dimensional structure'' phase transitions are determined at points $\rho(B_{crit1})=\varepsilon_0$ and $\Phi(B_{crit2})=\xi_0$.  

Similar reasoning is also correct for studying the dynamics of ideal models after the limited amount of time $\tau_{max}$. In the presence of the linear friction, the oscillation amplitude decays exponentially and takes a nonzero value at any $\tau\neq\infty$. We further consider the values of  $\varepsilon_{0}$ and $\xi_{0}$ in the equations of motion~(\ref{(eq:MotionOX)})--(\ref{(eq:MotionOY)}) for $\tau_{max}=T$.

\begin{figure}[h!]
	\begin{minipage}[ht]{\linewidth}
		\center{\includegraphics[width=\linewidth]{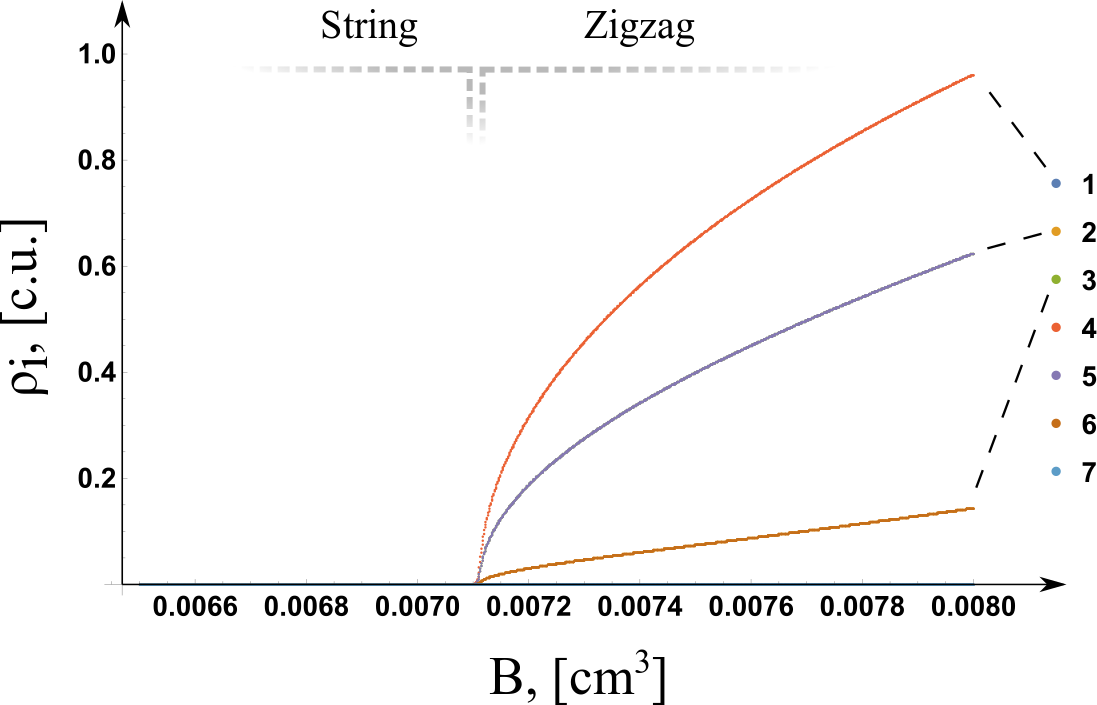}\\(a)}
	\end{minipage}
	\hfill
	\begin{minipage}[ht]{\linewidth}
		\center{\includegraphics[width=\linewidth]{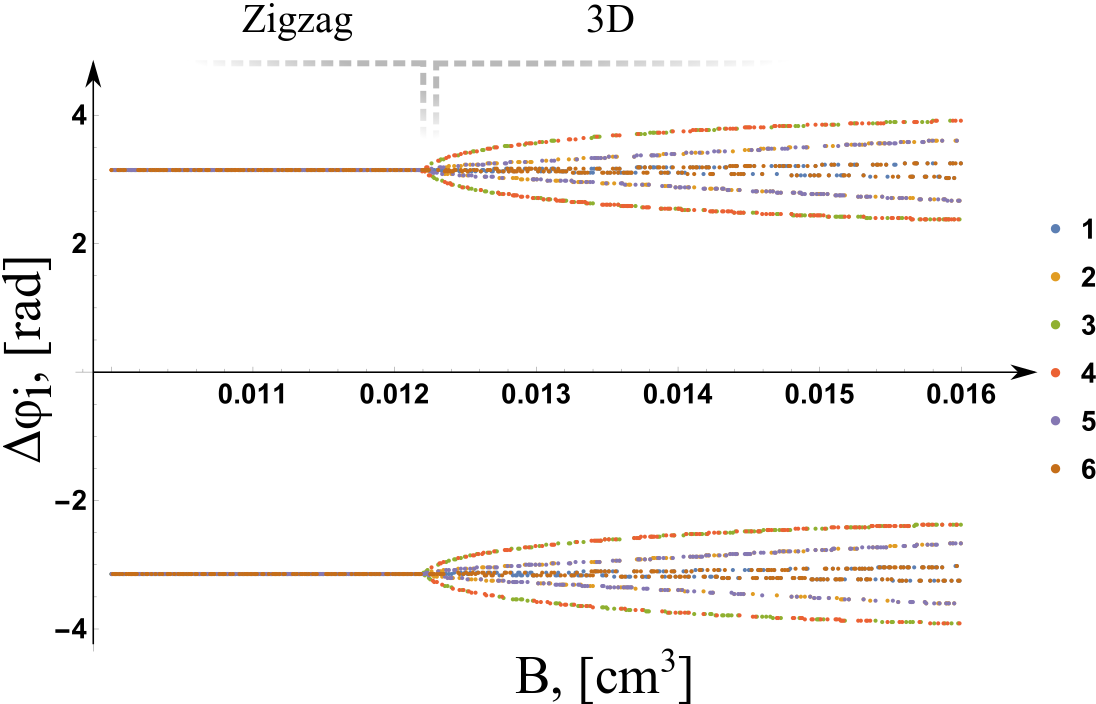}\\(b)}
	\end{minipage}
	\caption{Functions of geometrical parameters in the case of $N=7$ ions and $A=1$ a) the cross size $\rho_{i}(B)$; b) the relative angle between neighboring particles $\Delta \phi_{i}(B)$}
	\label{img:Rho(i)Phi(i)}
\end{figure}

Figure~\ref{img:Rho(i)Phi(i)} shows functions $\rho_{i}(B)$ and $\Delta \phi_{i}(B)$ for the system of $N=7$ ions when $A=1$ (Eq.~(\ref{(eq:MotionOX)})--(\ref{(eq:MotionOY)})). We calculate $\rho_{i}(B)$ and $\Delta \phi_{i}(B)$ for ions that have ``neighboring'' $z$ coordinates. The step $\delta B$ equals $10^{-3}$. The linear structure forms when $B$ is less than $B\approx0.007$. In this case, $\rho_i$ are low, and $\Delta\phi_i$ are undetermined. The cross size of the crystal increases substantially when the value of $B$ tends to $B_1$(Fig.~\ref{img:Rho(i)Phi(i)}a), and $\Delta \phi_{i}$ take on values $\pm\pi$ (Fig.~\ref{img:Rho(i)Phi(i)}b). Note that one can see only four branches of $\rho_i$ in Fig.~\ref{img:Rho(i)Phi(i)}a since zigzag has the central symmetry. When the value of $B$ crosses a $B_2\approx0.012$ point, $\rho_i$ continue to increase, and $\Delta\phi_i$ belong to $\left[0;2 \pi \right]$. Thus, one can observe phase transitions in terms~(\ref{eq:firstcond})--(\ref{eq:thirdcond}) near the points $B_1\approx0.007$ and $B_2\approx0.012$.

We intentionally say ``near the points $B_1$ and $B_2$'' and do not call $B_1$ and $B_2$ the critical points, as $\rho(B)$ and $\Phi(B)$ have no cusps at the points of phase transitions when $T$ is limited. According to the definitions~(\ref{eq:firstcond}--\ref{eq:thirdcond}), $\varepsilon$ tends to $\varepsilon_0$ at the point of the ``linear chain-zigzag'' phase transition only in the event of $T\to\infty$. In other words, the determination accuracy of the points of phase transitions is reliant on the limited period of time $T$. The values $B_{1}\to B_{crit1}$ and $B_2\to B_{crit2}$ under the condition $T\to\infty$.

\begin{figure}[ht!]
	\center{\includegraphics[width=\linewidth]{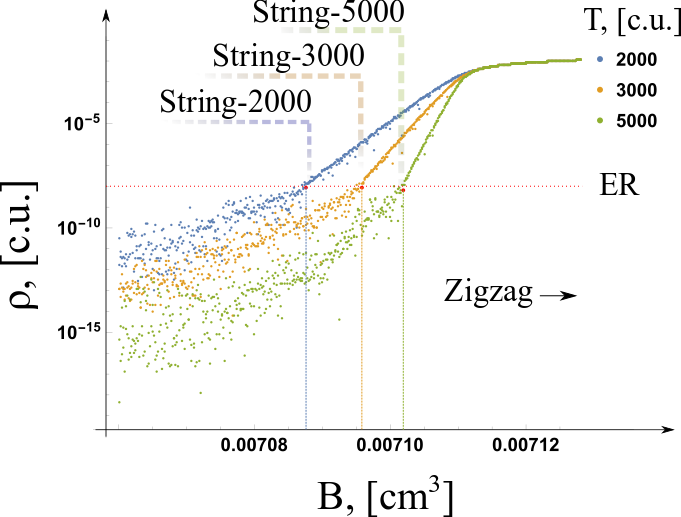}}
	\caption{Numerical calculation of the cross size function $\rho(B)$ in the case of $N=7$, $A=1$ near the point of the phase transition ``linear chain--zigzag''. The blue line corresponds to $T=2000$, orange line corresponds to $T=3000$, green line corresponds to $T=5000$. Error Rate~(ER) is an error order of the calculation. Red dots correspond to the values $\rho(B)=ER=10^{-8}$ for $T=2000$, $3000$, $5000$}
	\label{img:NumericCalcVaresilon0}
	\end{figure}

For clarity, we consider the numerical calculation of $\rho(B)$ in the case of $N=7$, $A=1$ near the point of the ``linear chain-zigzag'' phase transition ($B_{1}$) for dimensionless $T=2000, 3000, 5000$, Fig.~\ref{img:NumericCalcVaresilon0}. The graph is on a logarithmic scale.

The function $\rho(B)$ has no explicit cusp near the point $B_1$ on a finite time interval. According to the definition~(\ref{eq:firstcond}), the Coulomb crystal is linear when $\rho\to0$. In terms of numerical calculation, the minimal value of $\rho$ is limited by an accuracy rate. The red line in~Figure~\ref{img:NumericCalcVaresilon0} represents the Error Rate (ER). Thus, $B_1$ is limited by the maximum value $B$ at the point $\rho(B)=ER$, and $\rho<ER$ is true for all $B<B_{1}$ in consideration of the monotonic increase of the function $\rho(B)$. Nonetheless, the value of $B_1$ can be significantly different for various $T$ and $B_{1}\to B_{crit}$ for $T\to\infty$. The values of $B_1$ for $T=2000,3000,5000$ are shown in Figure~\ref{img:NumericCalcVaresilon0}.

\begin{figure}[ht!]
\begin{minipage}[ht]{\linewidth}
		\center{\includegraphics[width=\linewidth]{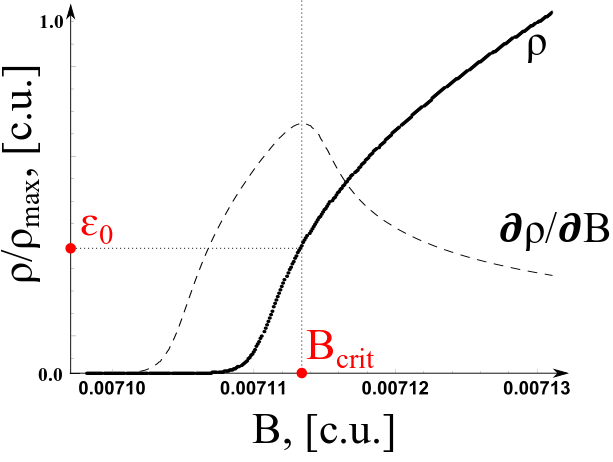}\\(a)}
	\end{minipage}
	\hfill
	\begin{minipage}[ht]{\linewidth}
		\center{\includegraphics[width=\linewidth]{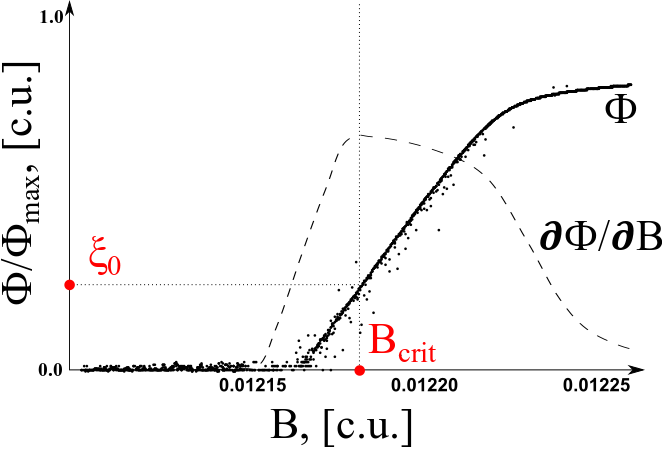}\\(b)}
	\end{minipage}
    \caption{The numerical calculation of the geometrical parameters functions a) the cross size $\rho(B)$ (solid line) and  the first derivative $\partial{\rho}/\partial{B}$ (dashed line); b) the metric $\Phi(B)$ (solid line) and the first derivative $\partial{\Phi}/\partial{B}$ (dashed line)} 
    \label{img:RHOandPHI}
\end{figure}

On the other hand, one can consider a dramatic increase of the function $\rho(B)$ when $B\to B_{crit}$. Moreover, the value of the cross size at which the derivative ${(\partial \rho / \partial B)}$ has maximum \textbf{is independent from $T$}. From this perspective, the point of the phase transition can be defined as the value of $B$ at which ${(\partial \rho / \partial B)}$ has the maximum and consequently $\rho(B)$ has the maximum growth rate. As shown in Figure~\ref{img:RHOandPHI}a, the function $\rho(B)$ monotonically increases in the range of $B\approx 0.007$ to $B\approx B_1$ but the crystal is still linear in terms (\ref{eq:firstcond})--(\ref{eq:thirdcond}) and the phase transition does not occur here. The growth of $\rho(B)$ in this range represents the fact that the phase transition is not instantaneous. Therefore, one can observe the phase transition ``linear chain--zigzag'' at the point $B_{1}=0.007113$, $\rho(B_{1})=\varepsilon_0=0.0299$ in~Figure~\ref{img:RHOandPHI}a. Note that this definition is consistent with ($\ref{eq:firstcond}$): the growth rate at the point $\varepsilon_0$ tends to infinity when $T\to \infty$. One can find the point $B_{2}=0.01218$ and the value $\xi_0$ in the similar way by the analysis of functions $\Phi(B)$ and $\partial{\Phi}/\partial{B}$ (Figure~\ref{img:RHOandPHI}b).
	
\section{Conclusion}

In the present work we have proposed a new method to determine phase transitions in Coulomb crystals in a linear ion trap. This method is based on the studying of the geometrical parameters of the ion crystal -- the cross size $\rho(B)$ and the metric $\Phi(B)$. We have considered a model where ions positions along the $z$ axis are fixed. In terms of this model, the linear structure can be characterized with $\rho(B)\leq\varepsilon$. The function $\rho(B)$ has a cusp at the point $B_1$, which is the point of the ``linear structure--zigzag'' phase transition. The zigzag structure is described by two parameters $\rho$ and $\Phi(B)$, and $\rho>\varepsilon$, $\Phi(B)\leq\xi$. The function $\Phi(B)$ has a cusp at the point $B_2$ which is the point of the ``zigzag--three-dimensional structure'' phase transition. In the case of a three-dimensional structure, $\rho>\varepsilon,\Phi(B)>\xi$. Hence, functions $\partial{\rho}/\partial{B}|_{B=B_1}$ and $\partial{\Phi}/\partial{B}|_{B=B_2}$ have asymptotes at the points of the ``linear structure--zigzag'' ($B_1$) and ``zigzag--three-dimensional structure'' ($B_2$) phase transitions, respectively. The remarkable fact is that $\varepsilon$ and $\xi$ are limited from below by nonzero $\varepsilon_0$ and $\xi_0$, when one considers a system for a limited period of time $T$. Moreover, the determination accuracy of $\varepsilon_0$, $\xi_0$ and, hence, the points of phase transitions, is reliant on the limited period of time $T$: the more the time $T$, the closer the value of $B$ to the true value $B_{crit}$.

The method proposed is necessary for the correct experimental data analysis and determination of points of phase transitions in real systems. For instance, one can describe theoretically unusual structures such as radial two-dimensional ion crystals that were experimentally observed recently~\cite{d2020radial}. Moreover, these criteria allow to discover previously unknown structures. 

\nocite{*}
\bibliography{aipsamp}

\end{document}